# Overaging with stress in polymer glasses? Faster segmental dynamics despite larger yield stress!

Masoud Razavi, Enran Xing, M.D. Ediger*

Department of Chemistry, University of Wisconsin–Madison, Madison, Wisconsin 53706, United States

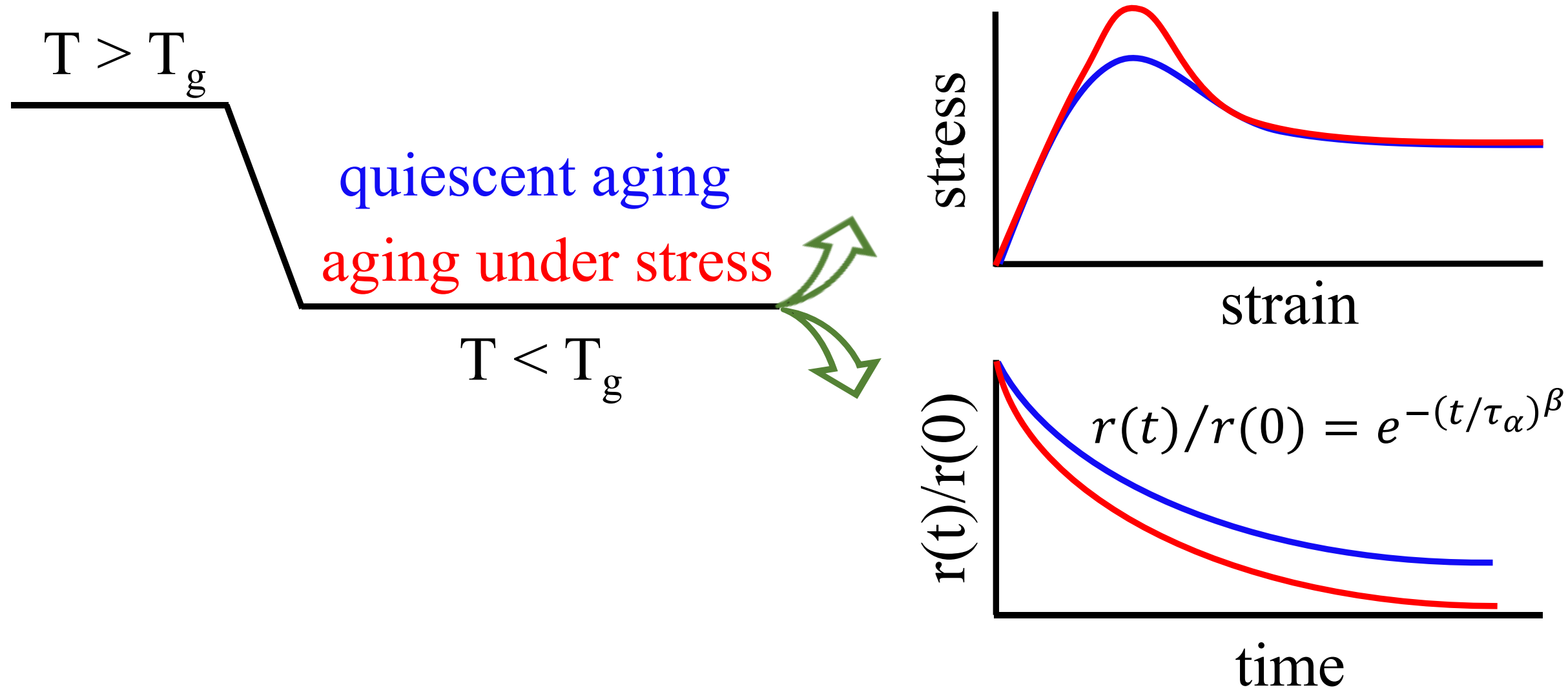

## ABSTRACT

It is well known that physical aging of polymer glasses increases their yield stress and affects their failure behavior. Studies indicate that application of moderate levels of stress during aging results in higher yield stress compared to aging in the absence of stress (quiescent aging). This has been interpreted to indicate that stress accelerates physical aging, and has been described as overaging. In this study, we age PMMA glasses under stress, and carry out direct measurement of segmental dynamics during and after aging by using a probe reorientation technique. We observe that samples aged under stress, despite having higher yield stress, have faster segmental dynamics after stress release than quiescently aged samples. This contradicts the overaging interpretation, for the range of conditions explored here. Our results indicate that yield stress is not a simple function of structural relaxation time and theoretical models based on this understanding need to be revised.

* Corresponding author email: ediger@chem.wisc.edu

## INTRODUCTION

Glassy materials are structurally disordered and thermodynamically non-equilibrium. Having long structural relaxation time at temperatures lower than the glass transition temperature makes these materials mechanically strong and suitable for many applications.[1] Processability of these materials is easier compared to their crystalline counterparts. In addition, because of the absence of grain boundaries between structural units, glasses provide an optical transparency advantage over polycrystalline materials. However, due to their non-equilibrium thermodynamic nature, the mechanical properties of glasses such as modulus, yield stress and creep compliance evolve over time.[2] This process of thermodynamic evolution toward the equilibrium state is known as structural recovery or physical aging.[3-6] From an application and design standpoint, it is very important to be able to predict the mechanical properties of a material at any time over its service life.[7-8] Hence, there have been extensive experimental[9], simulation[10] and modeling[11-12] studies on the correlation between structural, dynamic, and thermodynamic parameters and mechanical properties, as well as dependence of those properties on aging time, temperature, and pressure[13-14].

The potential energy landscape (PEL)[15] formulism is a successful approach that describes the thermodynamic and configurational changes that occur during physical aging. One consequence of physical aging on the dynamics of a glass is an increase of the structural relaxation time ($\tau_\alpha$). At a constant pressure, the rate of physical aging and magnitude of changes in the thermodynamic properties (e.g., free energy) or the dynamics (e.g., structural relaxation time) of a glassy material depend on the aging temperature. Aging is facilitated at higher temperatures where the mobility is high. The opposite of aging is rejuvenation, a process that moves the system upward in the PEL and results in accelerated dynamics.[16-19] It is intuitive that application of a temperature increase or external stimuli such as deformation[20] or radiation[21] can rejuvenate a glass. For example, the low energy packing arrangements achieved during a long aging experiment can be quickly destroyed by deforming the sample past yield.

Less intuitively, a process opposite to mechanical rejuvenation ("overaging") has sometimes been observed for glasses, both in experiment[22-23] and simulation[24-26]. Under some conditions, a small amount of mechanical excitation (e.g., repeated application of strains smaller than the yield strain) or irradiation can accelerate the movement of a system downward in the PEL and thus accelerate aging, an effect which termed as overaging. In these studies, overaging was quantified by measuring thermodynamic quantities such as potential energy and enthalpy. These observations are in analogy to accelerated dynamics of colloids[27] and the accelerated rate of compaction of granular systems subjected to the external agitation such as tapping or shaking[28-29]. At a qualitative level, overaging can be understood as follows: the mechanical deformation accelerates mobility in the glass (by lowering some energy barriers)[30] and the system is able to use this mobility to move down the PEL more rapidly.

The occurrence of overaging in experiments on polymeric glasses appears to be a controversial topic. Early experiments by Sternstein and Ho compared the aging of PMMA glasses under stress to quiescently aged glasses.[31] They observed significant changes in stress relaxation during repeated applications of stress (at levels below the yield stress). They interpreted these observations as the "mechanical enhancement of physical aging". This interpretation was later criticized by Struik,[32] who studied the linear and non-linear creep response of PMMA and PVC samples. Struik argued that the stress relaxation data of Sternstein "suffers from nonlinear effects", and concluded that, after a nonlinear deformation, the age of different samples must be compared in the linear viscoelastic region. McKenna and Zapas[33] took this argument another step further. They concluded that the analysis of the linear viscoelastic data obtained from superposing small strains on large strains was ambiguous, and that data should be analyzed in the context of a suitable nonlinear model[33]. In more recent studies on polycarbonate glasses, it has been observed by Govaert and coworkers[11-12] that the presence of stress during aging can increase the yield stress when the sample is subsequently subjected to uniaxial tensile deformation. Similarly, a larger yield stress as a result of aging under compressive deformation has been reported for PMMA glasses by Nanzai and coworkers[34]. Since the yield stress increases during quiescent aging, the more rapid increase in the yield stress than occurs during aging under stress can be interpreted as evidence of overaging. In the work of Govaert and coworkers, annealing time-stress superposition has been implemented to convert the effect of stress to the prolonged annealing time. In this approach, the presence of stress during aging is equal to aging for longer times in the absence of stress. Based upon this view, a constitutive model containing an aging term to describe the thermomechanical history of sample has been proposed,[11-12] and this model successfully predicts the yield stress and failure behavior of aged polymer glasses.[11-12] Motivated by this work, the impact of stress on aging was investigated by Lyulin and Michels in molecular dynamics simulations of quenched polycarbonate and polystyrene;[26] they observed that aging under stress increased the yield stress, and overaging was confirmed (as quantified by changes in the system energy). On the other hand, molecular dynamics simulations by Liu and Rottler[19] reached the opposite conclusion. While they found an increased yield stress for the sample aged under stress, they found no change in the rate of physical aging (as quantified by changes in the system energy); in this study, the increased yield stress was attributed to bond orientation rather than overaging.  In addition, there are several models of polymer glass deformation that do not allow for overaging;[35-38] it is an open question whether these models predict an increased yield stress for samples aged under stress[37-38].

Most of the previous experimental approaches for assessing the age of polymer glasses after aging have been indirect, based upon mechanical methods such as measurement of yield stress, creep response or stress relaxation behavior. As an alternate approach, here we utilize photobleaching measurements of the segmental dynamics to characterize the age of poly(methyl methacrylate) (PMMA) glasses. Since aged samples are known to have slower segmental dynamics, this provides quite a direct test of the overaging interpretation. We compare the segmental dynamics in two conditions: i) isothermal physical aging in absence of stress (quiescent aging) and ii) isothermal physical aging during and after the application of a constant stress. The aging protocols used here and their mechanical consequences are similar to the study on polycarbonates[11-12]. In agreement with Govaert's work[11-12], we observe larger yield stresses for the samples aged under stress. In contrast with Govaert's interpretation, segmental relaxation times of these PMMA samples are always smaller than (or the same as) quiescently aged samples. The effect of stress on segmental relaxation time is found to be similar to the theoretical calculations by Chen and Schweizer on PMMA[39]. Results of our study indicate that the increased yield stress of a polymer glass, due to aging under stress, is not necessarily an indication of overaging and our understanding of the relationship between segmental relaxation time and yield stress of polymer glasses, and theoretical models based on this understanding, need to be revised.

## EXPERIMENTAL

**Sample Preparation.** Lightly crosslinked PMMA was synthesized and used as a model polymer glass in our study. It is anticipated that crosslinking does not significantly affect segmental dynamics.[40] Lightly crosslinked PMMA embedded with a low concentration of probe molecules was synthesized by free radical polymerization as follows. Initially a solution of methyl methacrylate (MMA, monomer), ethylene glycol dimethacrylate (EGDMA, crosslinker) with concentration = 1.3 wt%, and N,N′-dipentyl-3,4,9,10-perylenedicarboximide (DCCP, probe molecule) with concentration = $10^{-6}$ M was prepared. Then, benzoyl peroxide (BPO, initiator) at a concentration = 0.1 wt% was added into the prepared solution and the mixture was heated and kept at 343 K for approximately 30 mins. MMA, EGDMA and DCCP were obtained from Sigma-Aldrich, and BPO was acquired from Polysciences, Inc.. The mixture was spread onto a rectangular microscope slide prior to gelation when it became relatively thick. Strips of aluminum foil as spacers were placed at the corners of the slides.  A second slide was placed on top, and the sandwich was clamped together and then transferred into a vacuum oven. Polymerization was continued for 24 hr at 343 K and then another 24 hr at 393 K. After completion of polymerization the polymer film was removed from the microscope slides by sonication. The glass transition temperature ($T_g$) of polymer was measured to be 400 K using differential scanning calorimetry (DSC). A scaled-down ASTM D1708-10 die cutter was used to cut dogbone samples out of the polymer film. The sample length and width in the mid-section were respectively equal to 13 mm and 2 mm. The thickness of a typical sample ranged between 30 to 40 μm, with the thinnest region in the middle where the optical experiments were performed. All the mechanical and optical measurements shown in the main text were performed on one sample. Thermal annealing above $T_g$ (ca. $T_g + 20$) for sufficient time (ca. 1 hr) removed the prior deformation and thermal history and allowed repeated measurements on the same sample. The repeatability of the measurements was also confirmed by performing similar measurements on different samples.

**Mechanical measurement.** The dogbone sample was clamped on the stretching grips and loaded into the temperature-controlled deformation cell. Prior to measurements, the sample was annealed at $T_g + 20$ K for 1 hr. Then it was cooled to the measurement/annealing temperature with a cooling rate of 2 K/min. All the mechanical deformations beyond yield were performed at 380 K since PMMA is brittle at lower temperatures. For some annealing protocols, we annealed the sample at lower temperatures, e.g. 360 K, and then followed with mechanical testing at 380 K. For this case, the heating rate to increase the temperature from lower temperatures to 380 K was also 2 K/min. Tensile deformation was carried out at 380 K with the global deformation rate of 0.01 $s^{-1}$ to identify the yield stress of samples after different thermal and mechanical annealing conditions.  For samples aged under stress, the constant strain rate deformation was performed about 2 minutes after the stress was set to zero. During this time, the sample retracted slightly (roughly 1%, details can be found in Table 1). We report dynamic yield stress values here, i.e. the yield stress is measured as the peak of stress-strain curve. While other yield stress quantities have been utilized,[41] this is the most common approach and it is the approach used in ref. 11.

**Probe reorientation measurement.** The segmental dynamics of the sample during aging (in the presence and absence of stress) were measured using the photobleaching technique[42]. In this method, a linearly polarized laser beam preferentially photobleaches the DPPC probe molecules with transition dipoles aligned to the polarization direction of the laser beam. This results in an anisotropic distribution of orientation of unbleached probes that is detected by a weak circularly polarized laser beam that induces fluorescence. Initially, the fluorescence intensity with polarization parallel to the photobleaching polarization is lower than that in the perpendicular direction. Due to reorientation of the probes over time, the anisotropy decreases, and the two intensities converge. The anisotropy decay function r(t) can be defined based on these two intensities[42]. This function decays over time and can be described by the Kohlrausch−Williams−Watts (KWW) function:

$$r(t) = r(0)e^{-(t/\tau)^{\beta}}$$

Here, r(0) is the initial anisotropy, τ is the probe reorientation relaxation time and β characterizes the nonexponentiality of the decay. It has been shown that reorientation of DCCP can accurately characterize segmental dynamics of PMMA.[43] Previous studies indicated that the β value for probe reorientation in PMMA glass is 0.31 in the absence of deformation.[42] For the anisotropy decay data recorded in the absence of deformation, we constrained the fitting to the KWW function by fixing β = 0.31, while β was freely fitted for samples under stress. In the latter case this parameter ranged between 0.28 to 0.31. Constraining β to be constant at 0.31 does not result in a meaningful difference in the fitted relaxation times compared to leaving it free during fitting.

## RESULTS

The goal of this study is to look for evidence of overaging in lightly-crosslinked PMMA glasses. We compare the behavior of samples aging with and without an external stress. To allow a comparison with previous studies,[11-12] we perform constant strain rate deformations after aging, in order to determine the yield stress. To provide further insight, we use a photobleaching technique to characterize the segmental dynamics of these glasses during aging (during and after the imposition of an external stress). Figure 1 introduces the key elements of the experiment. Panel a shows the probe molecule present at low concentration (~$10^{-6}$ M) in the polymer glass; the average probe reorientation time characterizes the segmental dynamics. Panel b shows the sample cell that allows simultaneous mechanical and probe reorientation measurements. Panel c shows typical yield stress measurements and panel d shows typical probe reorientation data.

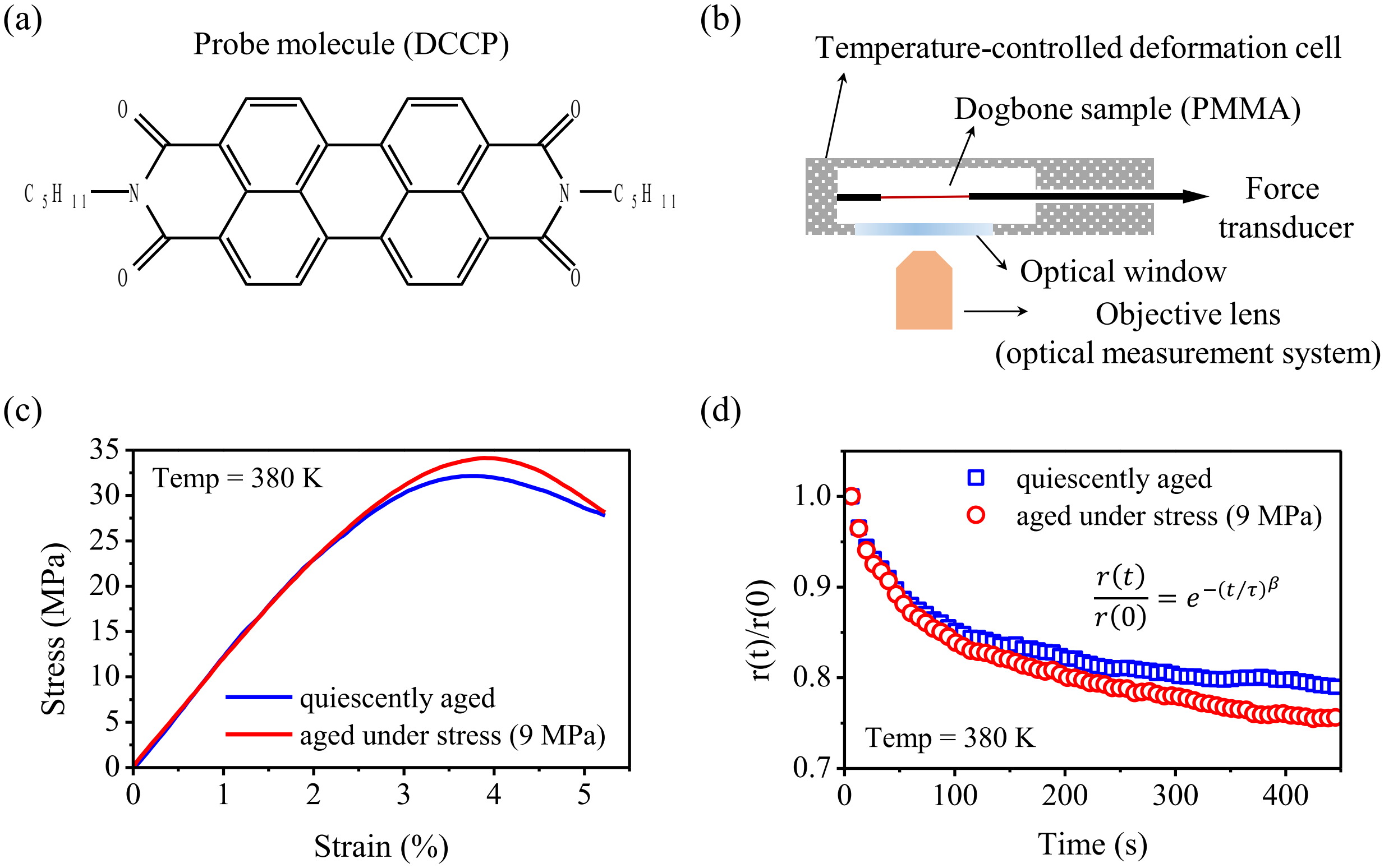

**Figure 1.** (a) Chemical structure of DCCP; the transition dipole is parallel to the long axis, (b) Schematic diagram of the temperature-controlled mechanical deformation cell and the optical measurement system. For detailed information see ref. 42. (c) Stress-strain curves for a quiescently aged sample and a sample aged under stress of 9 MPa, indicating a higher yield stress for the sample aged under stress. Strain on the x-axis was calculated based on the sample length just prior to the constant strain rate experiment. (d) Anisotropy decay curves for a quiescently aged sample and a sample aged under stress of 9 MPa (but after the stress was released), indicating faster segmental dynamics as a result of aging under stress. The sample aged under stress has faster segmental dynamics despite its higher yield stress.

Prior to measurements of probe reorientation, we performed a series of experiments to identify conditions where aging under stress led to an increase in the yield stress, as this is the key observation of ref. 11. For the thermal history utilized here (upper panel in Figure 2a), the yield stress of the sample immediately after reaching the test temperature (380 K) was equal to 27.7 MPa. Aging of this sample in absence of stress, i.e. quiescent aging, for 72000 s increases the yield stress to 32.2 MPa, which is 16 % more than the yield stress at t = 0. Aging under stress of 9 MPa according to the protocol shown in the lower panel of Figure 2b for the same time (72000 s) results in a yield stress of 34.1 MPa, 6% higher than the yield stress of quiescently aged sample. Constant strain-rate data for these two aged samples are compared in Figure 1c. The aging protocol shown in Figure 2a produces mechanical data that could be interpreted as indicating that the sample aged under stress is overaged compared to the quiescently aged sample. In other mechanical experiments we found qualitatively similar results for aging under stresses in the range of 6-12 MPa; aging under stresses higher than 12 MPa reduces the yield stress (rejuvenation). Since aging at 9 MPa produced the greatest increase in yield stress (relative to quiescent aging), we adopted this condition for optical measurements of probe reorientation.

Figure 2b compares the segmental relaxation times measured during quiescent aging, and during and after aging under 9 MPa stress. For each aging condition, we started a series of photobleaching experiments around 1000 s after reaching the aging temperature (380 K); each photobleaching experiment yields a single point in Figure 2b. Quiescent aging is well described by the power law with the exponent of ≈ 0.53. Segmental relaxation for the glass under stress is about a factor of 2 faster than during quiescent aging. Qualitatively, this result is expected as many previous experiments have shown that stress accelerates segmental motion.[3, 42-45] The empty red circles show the segmental relaxation time immediately after stress removal. We continue to observe faster dynamics after the stress is released, relative to quiescent aging, and this effect persists for at least 30,000 seconds (at which point the experiment was ended). Figure 1d shows examples of anisotropy decay curves associated with the two aging conditions.

Figure 2 directly addresses the overaging hypothesis. While the increased yield stress caused by aging under stress could be interpreted as overaging, the segmental dynamics indicate the opposite. Once the stress is released, the segmental dynamics are faster than the quiescently aged glass, consistent with the view that a *younger* glass is produced by aging under stress.

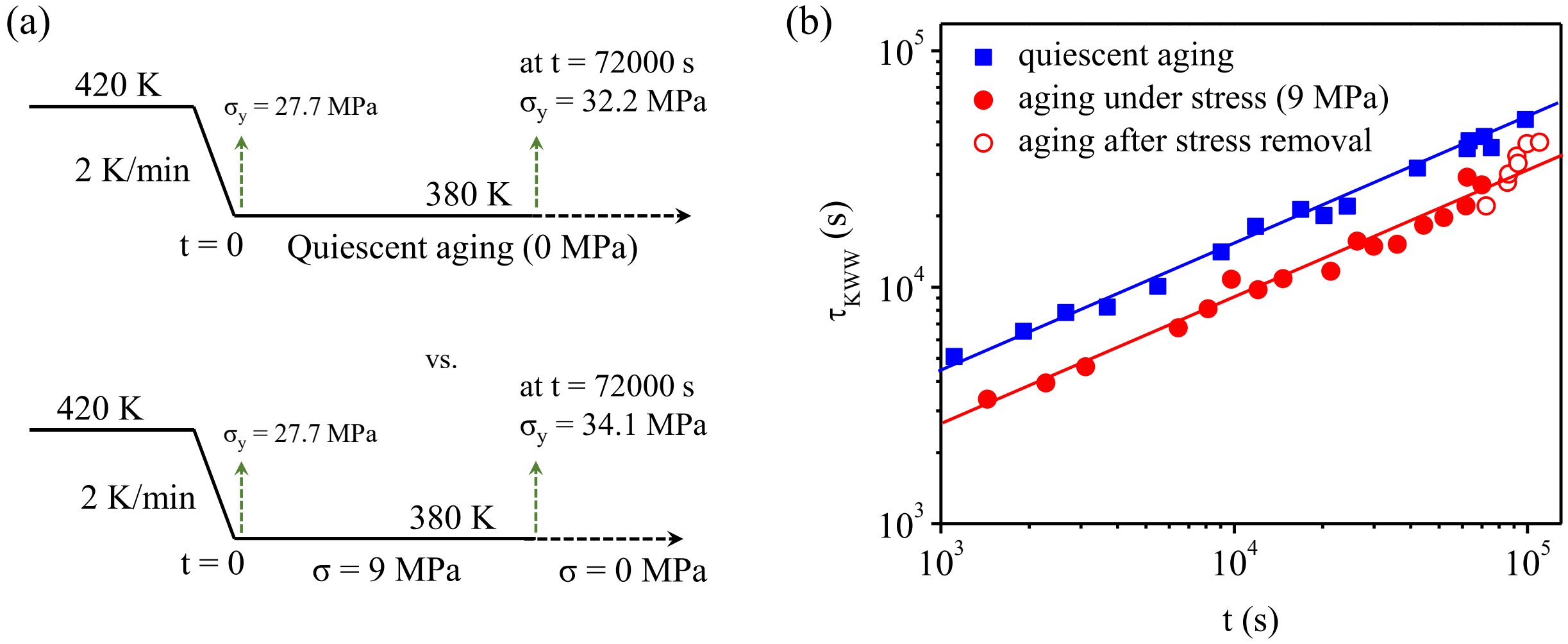

**Figure 2.** (a) Temperature profile and stress condition during quiescent aging (upper panel) and aging under stress (lower panel). During aging under stress, a stress of 9 MPa was applied into the sample for 72000 s. The green dashed arrows show the yield stress recorded in a constant strain rate deformation. The constant strain rate deformation was performed about 2 minutes after the stress was set to zero. (b) Comparison of segmental relaxation times for the two aging conditions shown in panel (a).

One could look at the dynamic data in Figure 2b and envision that, at much longer aging times, the glass aged under stress might have slower dynamics than the quiescently aged glass, and if so, some version of the overaging interpretation might be sustained. To test this idea, we performed two further experiments. Figure 3a shows a very similar experiment to Figure 2, with the difference being that the stress was removed at 18000 s rather than at 72000 s. The empty red circles are the data collected after stress removal and the dashed line is the extension of the continuous red line. Segmental mobility after stress removal decreases and appears to be rejoining to the aging trajectory of the quiescently aged sample but clearly does not overage, even after waiting 7 times longer than the time used for aging under stress. We emphasize that presence or absence of such recovery would not change our observation and conclusion regarding yield stress values in Figure 2. The yield stress of the sample aged under stress at the moment when it has faster segmental dynamics (dynamically younger) is higher than the yield stress of sample with slower segmental dynamics (dynamically older sample).

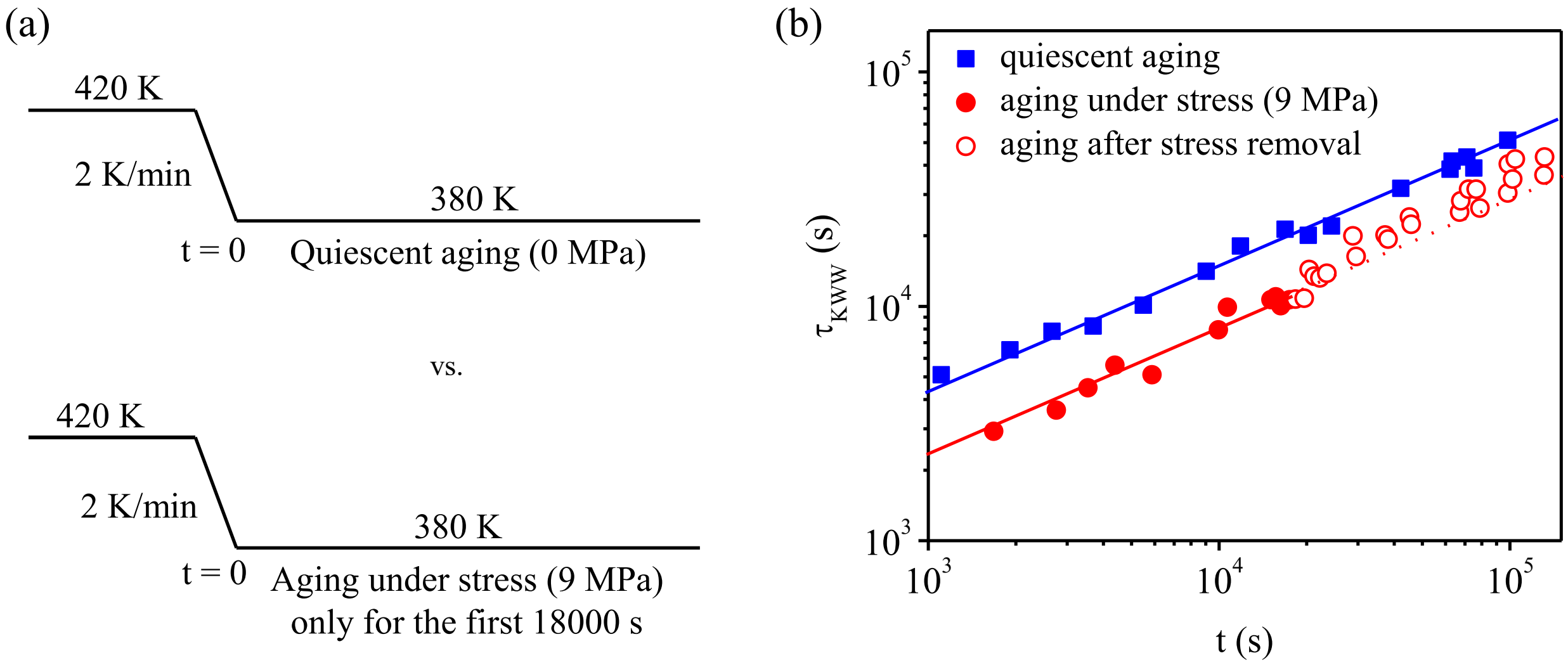

**Figure 3.** (a) Temperature profile and stress condition during quiescent aging (upper panel) and aging under stress for 18000 s (lower panel). (b) Comparison of the segmental relaxation times for samples with the two aging conditions.

A final set of experiments were performed to allow an even greater opportunity for recovery after aging under stress. Figure 4a shows two aging protocols. The aging temperature was 360 K and samples were aged for 72000 s. A stress of 15 MPa was used for one of the samples. Dynamic data were collected both at 360 K and also after the temperature was raised to 380 K. Yield stresses were recorded upon reaching 380 K. The yield stress of the quiescently aged sample was 29.7 MPa. The value for the sample aged under stress was 3% higher and equal to 30.5 MPa. The dynamic measurements at 360 K in Figure 4b shows accelerated mobility in presence of stress, as expected from previous work.[44] Similar to Figures 2 and 3, the sample aged under stress remains younger after stress removal. Again in this experiment, while the yield stress could be interpreted in terms of overaging, the segmental dynamics contradict this interpretation. After sufficient time at 380 K, it appears that the impact of aging under stress disappears, as the dynamics of the samples follow the same time course (note the expanded inset plot). The faster dynamics recovery in this case (aging at 360 K) compared to the previous case (aging at 380 K) may be due to the smaller ratio of the aging time ($t_{aging}$) to the segmental relaxation time $\tau_\alpha$ at 360 K versus at 380 K. In other words, the rate of dynamic recovery after stress removal might be proportional to $t_{aging}/\tau_\alpha$. Regardless of the rate of dynamic recovery, dynamic data of the sample aged under stress at longer times overlaps with the aging trajectory of the quiescently aged sample, without exceeding them, and thus provide no support for the overaging interpretation. A similar feature, i.e. recovery of dynamic data to that of the quiescent aging condition, has been observed in a previous study when the sample was subjected to repetitive cyclic deformations[46]. Our experimental observations regarding effect of stress on aging behavior of PMMA are in close agreement with the theoretical calculations of Chen and Schweizer, that are discussed in more detail in the discussion session[39].

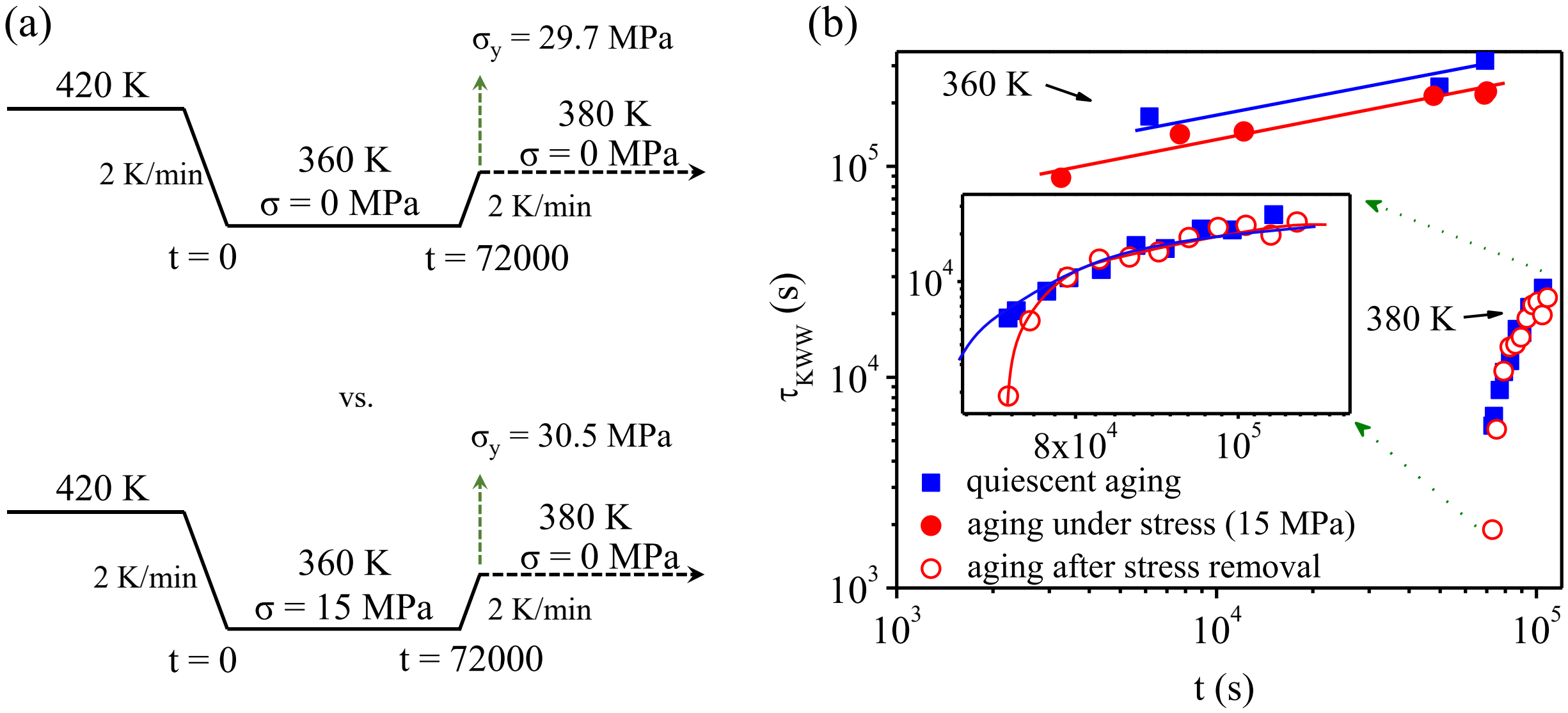

**Figure 4.** (a) Temperature profile and stress condition during quiescent aging (upper panel) and aging under stress of 15 MPa (lower panel). Stress was applied at 360 K while the yield stress was measured at 380 K. The green dashed arrows show the yield stress recorded in a constant strain rate deformation. The constant strain rate deformation was performed about 2 minutes after the stress was set to zero. (b) Comparison of the segmental relaxation times of the samples with the two aging conditions. The inset presents an expanded version of the 380 K portion of the experiment and uses a linear time axis.

A more detailed evaluation of the results of aging at 380 K from the perspective of references 11 and 12 can be performed using Figure 5. The yield stresses for a series of quiescently aged samples at 380 K were measured. As shown, the yield stress increases linearly with the logarithm of aging time (filled blue squares), similar to the results in ref. 11. t = 0 corresponds to the start of aging at 380 K. The filled red circle shows the yield stress of the sample under stress of 9 MPa for 72000 s. Two comparisons can be made. First, for the specific aging time of 72000 s, as previously pointed out, the sample aged under stress has faster dynamics despite its higher yield stress. Second, according to the references 11 and 12, the effective age of a sample under stress would be equal to the age of the quiescently aged sample with the same yield stress. By extrapolation, Figure 5 shows that around 400000 s of quiescent aging would be required to achieve a yield stress of 34 MPa (the value achieve by aging under stress), but at this time the expected relaxation time would be near 104000 s, about 4 times larger than predicted by the proposal of references 11 and 12. This relaxation time was estimated by extrapolating the data in Figure 2b.

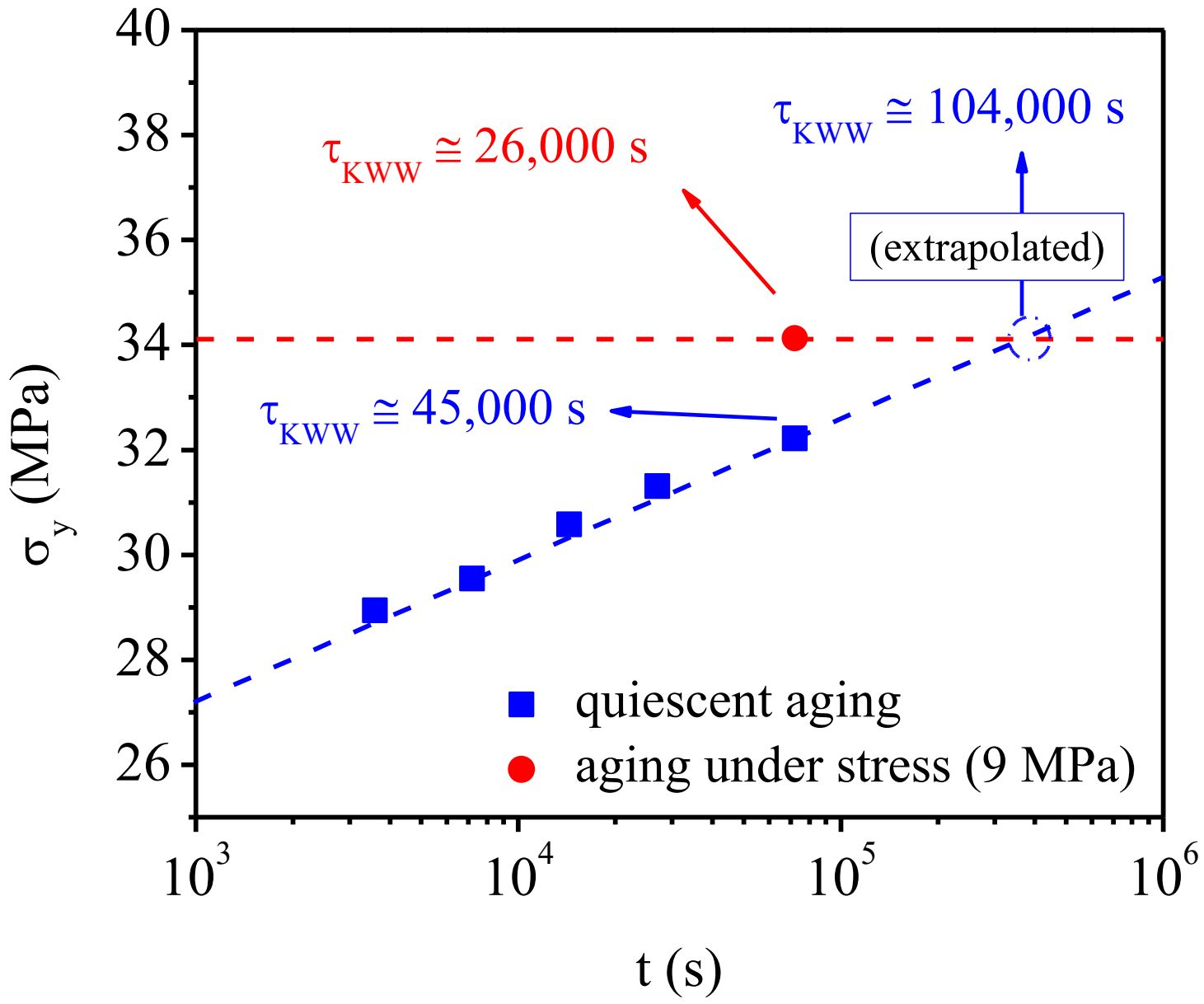

**Figure 5.** Evolution of yield stress during quiescent aging at 380 K with comparison to the yield stress of the sample aged under stress of 9 MPa. The extrapolated $\tau_{KWW}$ is obtained from the data in Figure 2b.

Measurement of the increase in the length of samples aged under stress might contain useful information regarding cause(s) of the enhanced yield stress in samples aged under stress. Table 1 shows the length increase of the samples aged at different conditions of temperature and stress for the aging time of 72000 s. The total creep is the percentage increase in sample length after 72000 s, just prior to stress removal. The permanent creep is the length increase, as measured about 2 minutes after stress release. We can also assess the impact of chain orientation through the *g* factor, details of which has been given in ref. 42. This factor depends upon the average alignment of the molecular probes along the deformation axis (a smaller *g* indicates greater alignment). In ref. 42, it was found that the *g* factor decreases 20% as the strain increases to 60 % and it drops by 40% for strains around 165 %. In the current study, for aging under stress of 9 MPa at 380 K, the *g* factor is almost constant; at the beginning of aging was 1.002 ± 0.005 and it dropped to 0.989 ± 0.004 at the end of 20 hr.

Table 1. Measurement of the increase in the sample length after aging for 72000 s at different aging conditions.

| Temp. (K) | Stress (MPa) | Total creep (%) | Permanent creep (%) |
|---|---|---|---|
| 380 | 9 | 5.8 | 5.1 |
| 360 | 15 | 2.6 | 1.5 |
| 340 | 20 | 2.1 | 0.8 |

**DISCUSSION**

In this work, we performed experiments to test the view that aging a polymer glass under stress "overages" the glass, i.e., the view that aging under stress for a given time is equivalent to quiescent aging for a much longer period of time. We find that, under these experimental conditions, the segmental dynamics of the polymer do not become longer than the quiescently aged sample as a result of aging under stress. Strikingly, immediately after the release of stress, segmental dynamics are faster than for the quiescently aged sample. This result contradicts the overaging hypothesis, for the range of conditions explored here. In this section, we discuss possible interpretations of our experimental results.

One possibility is that overaging does occur in the experiments reported here, but that the probe reorientation measurements are not reliably reporting the segmental dynamics of the polymer. It is true that our experiments directly measure the reorientation time of an ensemble of probe molecules and our statements about the segmental dynamics of the polymer are inferences. Past comparisons with probe-free methods of characterizing segmental dynamics have shown that the probe reorientation measurements are reliable in polymer melts[47-49] and, for linear deformations, in polymer glasses[42, 45]. For nonlinear deformation of polymer glasses, there are few experiments that directly report segmental dynamics, but the existing data is qualitatively consistent with probe reorientation measurements[50-51]. In addition, the experimental observations presented here agree with the theoretical calculations of Chen and Schweizer[39] using nonlinear Langevin equation (NLE)[52-53] on PMMA. They studied the effect of different stress levels on aging behavior of PMMA at $T_g$ - 10 and $T_g$ - 50. It was found that depending on the stress level bifurcation in the density fluctuation amplitude and relaxation time can take place. PMMA rejuvenates at higher stress values and ages under lower stress levels. However, the glass remains always younger than the quiescently aged sample (as indicated by the state variable). Similar to the experiments reported here, the rate of evolution of the segmental relaxation time, i.e. slope of $\tau_\alpha$ vs $t_{aging}$, of the sample aged under stress of 8.5 MPa at $T_g$ - 10 and that of aged under 14.5 MPa at $T_g$ - 50 are same as the quiescently aged samples at the corresponding temperatures. By increasing stress level, slope of $\tau_\alpha$ vs $t_{aging}$ reduces and finally becomes negative in the rejuvenation regime. Qualitatively similar effects were observed in aging of bead-spring polymer glasses under stress, in a molecular dynamic simulation study by Rottler[54]. The results [54] were found to be in agreement with theory[39] and both show rejuvenation and physical aging compete with each other in an additive manner. Therefore, in the NLE model and these simulations, aging under stress is possible but "overaging" under stress is not. In light of these comparisons, we find it quite unlikely that the probe reorientation measurements would be qualitatively in error in the experiments reported here.

A second possibility is that differences in experimental conditions for this work and ref. 11 explain why overaging is not observed here. Ref. 11 performed experiments on polycarbonate samples with $M_n \approx 10{,}000$ g/mol. Samples were aged under tension at 353 K (~ $T_g$ - 60 K) and then the yield stress was determined in tensile deformation at room temperature (~ 295 K). Ref. 11 reports an increase in the yield stress due to aging under stress and interprets this result as overaging. In the work reported here, cross-linked PMMA was utilized, and samples were aged under tension at 380 K ($T_g$ - 20 K) and 360 K ($T_g$ - 40 K). The yield stress was determined in tension at 380 K. It is possible that one of these differences between the two experiments allows overaging in polycarbonate even though we do not observe it in PMMA. However, as the key experimental signature of ref. 11 was reproduced here (increased yield stress for aging under stress), and the aging temperatures relative to $T_g$ are similar, we expect that our conclusion (no overaging) in PMMA also applies in polycarbonate. We point out that the model developed in ref. 11, based upon the assumption of overaging, makes useful predictions for time-to-failure for polycarbonate glasses aged under stress; at present, we have no explanation for this. The contradiction between our work and ref. 11 is deeper than the interpretation in terms of overaging. The model developed in ref. 11 predicts an increased viscosity as a result of aging under stress, which is connected with an increase in the segmental relaxation time. In contradiction to this prediction of the model of ref. 11, we observe a decreased segmental relaxation time following aging under stress.

Nanzai and coworkers[34] studied the properties of PMMA glasses aged under a constant compressive strain of 4%, and their work is quite relevant. Nanzai and coworkers reported faster evolution of the compressive yield stress with aging time in PMMA samples when the samples were aged under constant compressive strain. These authors also reported DSC thermograms of the samples aged with and without compressive strain, as a function of aging time. The quiescently aged samples show large changes in the DSC curves as a result of aging, consistent with the expected significant decrease in enthalpy. In contrast, the DSC curves for samples aged under compressive strain show almost no change, indicating essentially no enthalpy change. In our view, this result of Nanzai and coworkers is consistent with our own results. In each case, deformation-induced increases in the yield stress could be interpreted as overaging, but more direct information about the state of the glass (segmental relaxation times or enthalpy measurements) do not support this interpretation.

It is useful to explore some speculative ideas about why the model of ref. 11 does not correctly predict the shorter segmental relaxation times observed in our experiments. One possibility is that additional factors beyond the polymer age might influence the yield stress. In computer simulations, Liu and Rottler[19] found that the enhanced yield stress of the sample aged under stress correlated with the average orientation of the covalent bonds. We can test this idea for our experiments as follows. Using recent measurements of the strain hardening modulus of lightly crosslinked PMMA samples[55] prepared in the same manner as those used here, we can calculate the increase in yield stress expected as a result of chain orientation for the strains reported in Table 1. In all cases, the yield stress increase reported here is more than twice that expected on the basis of this simple calculation. This is consistent with the very small changes in the probe alignment (*g* factor values) reported in the results section. Thus, we conclude that chain orientation likely is a contributing factor but this is not the primary factor responsible for the increased yield stress values observed in our experiments for aging under stress.

Recent work by Medvedev and coworkers[56] provides another way to interpret the experimental results reported here. These authors criticize the "material time" assumption made by many models, including ref. 11. Qualitatively, the material time approach assumes that the non-linear aspects of polymer glass deformation are completely controlled by deformation-induced changes in the segmental relaxation time. In material time models, changes in the segmental relaxation time are controlled by deformation parameters like stress[57] and strain rate[36], or by state parameters such as configurational energy[58], local density fluctuations, etc. Ref. 56 reports experiments in which the segmental relaxation time is measured during a multi-step deformation. They conclude that the experimental results cannot be explained by standard models utilizing the material time assumption. Ref. 56 introduces a toy model that allows the elastic shear modulus to also respond to the deformation, and this model qualitatively reproduces the experiments. For this toy model, one does not expect a one-to-one correspondence between the yield stress and the segmental relaxation time. In this framework, an increase in the yield stress might be consistent with a decrease in the segmental relaxation time. Indeed, in this framework, the "age" of a polymer glass is not well-defined, as the age inferred from the relaxation time might be different than the age inferred from the modulus. NLE theory framework[37, 59] is an older approach that, similar to the toy model of Medvedev and coworkers[56], can potentially reproduce the higher yield stress of the samples aged under stress. As discussed in the preceding paragraphs, the NLE model does not allow overaging as it cannot predict a longer segmental relaxation time than that of the quiescently aged samples[39].

Many computer simulations of glasses have reported overaging as a result of deformation and one might ask why overaging occurs in simulations but not in the experiments reported here. In computer simulations, overaging is usually identified as a decrease of the inherent structure energy that occurs more rapidly during deformation than is observed from physical aging alone. Previous simulation work on polymer glasses has established a correlation between molecular mobility and inherent structure energy,[60] thus it is reasonable to expect that the mobility of overaged glasses would be lower than a quiescently aged system. One important difference between simulations and typical experiments on polymer glasses is the cooling rate used to prepare the glasses. Simulations commonly cool on the order of $10^{12}$ K/s[24-26, 61] a cooling rate more than 13 orders of magnitude faster than our experimental cooling rate. Recently, it has been established that the overaging effect is highly dependent on the extent of annealing in the glass, with more annealed glasses exhibiting less or no overaging[62-64]. It is possible that we did not observe overaging in our experiments because the glasses were cooled slowly and already reasonably annealed before the experiment began. If this explanation were to be correct, then one would not expect overaging in typical experiments on polymer glasses.

We conclude our discussion by briefly recounting the historical connection between yield stress and the "age" of amorphous materials. Prior to the work of ref. 11 and the interpretation of the higher yield stress as "accelerated aging under stress", Kramer carried out similar experiments on Nylon 6-10 monofilaments.[65] Kramer used "stress aging" to refer to the delayed yielding and neck propagation of partially drawn Nylon 6-10 monofilaments that had been previously subjected to stress values less than the stress used to propagate the neck. He interpreted this as a result of the formation of stress-induced microcrystals in the nominally amorphous domains of nylon.[65] Unlike ref. 11, the aging effect in Kramer's study was not interpreted as the physical aging of the glassy state. It is likely that "stress aging" and its conceptual connection to the yield stress was initially invoked in the metallurgical literature, where aging refers to the process of annealing of alloys; here heat treatment leads to nucleation and precipitation of a second phase out of uniform mixture of metals[66]. The consequence of this treatment is an increase in the strength and yield stress of alloys.[67] It has been observed that application of stress during heat treatment encourages the precipitation process and results in larger increase in the yield stress.[68-70] Recently it has been shown that yielding of high-$T_g$ semicrystalline polymers can occur with two yielding peaks, an initial yielding of the glassy phase followed by yielding of the semicrystalline phase.[71] In light of this, the observed behavior in Kramer's study is likely a glassy feature rather than some events in the crystalline phase and in this regard the governing physics would not be different than in polycarbonate[11-12] and PMMA. Moreover, in metallurgical studies, the faster development of yield stress with time in samples aged under stress compared to the normally treated samples could be due to the increased mobility under stress that can facilitate de-mixing and the precipitation process. In other words, even in the metallurgical studies the effect of stress is to increase mobility, as we report here for PMMA glasses.

## CONCLUSION

An accurate description of yield stress in polymer glasses is crucial from both fundamental and application points of view. To achieve more insight regarding the effect of physical aging on yield stress, we compared physical aging of PMMA glasses in two different conditions: quiescent aging and aging in the presence of stress. We observed that glasses aged under stress have higher yield stress and shorter segmental relaxation times. These results contradict the "overaging" interpretation in which aging under stress is viewed as equivalent to much longer aging in the absence of stress. We conclude that the yielding of polymer glasses is not solely determined by the segmental mobility, and there must be other parameter(s) to define the yield stress. One possible interpretation of these experiments is suggested by computer simulations[19] where a good correlation was observed between the increase in yield stress for polymer glasses aged under stress and an increased orientation of covalent bonds. Another more general interpretation is that the "material time" approach utilized in most modeling is inadequate[56], and that variables beyond the segmental relaxation time control the non-linear aspects of glass deformation[56, 59]. For laboratory polymer glasses, a question which remains unanswered is this: is there a regime where application of stress can dynamically accelerate physical aging and result in overaging, as confirmed by direct measurements of the sample energy or segmental mobility? Overaging under stress has been observed for *rapidly quenched* polymer glasses in molecular dynamics simulations[26]. However, the much slower cooling rates typical of laboratory polymer glasses may preclude the observation of overaging.

**ACKNOWLEDGEMENTS**

We thank the National Science Foundation (DMR-2002959) for support of this work. MDE thanks Leon Govaert for many helpful conversations.